\documentstyle[preprint,aps]{revtex}
\tightenlines
\begin{document}
\draft
\preprint{SUNY-FRE-98-01}
\title{Monopole Loop Suppression
and Loss of Confinement in Restricted Action
SU(2) Lattice Gauge Theory}
\author{M. Grady}
\address{Department of Physics\\ SUNY College at Fredonia\\ 
Fredonia NY 14063 USA}
\date{\today}
\maketitle
\begin{abstract}
The effect of restricting the plaquette ($1\times 1$ Wilson loop)
to be greater than a certain cutoff
is studied. The action considered is the standard 
Wilson action with
the addition of the plaquette restriction, 
which does not affect the continuum limit.
A deconfining phase transition occurs as the cutoff is raised, even
in the strong coupling limit.
Abelian-projected
monopoles in the maximal abelian gauge are strongly suppressed
by the action restriction.
Analysis of the steeply declining 
monopole loop distribution function indicates that for
cutoffs $c > 0.5$, large monopole loops which are any 
finite fraction of the 
lattice size do not exist in the infinite
lattice limit. This would seem to imply
the theory lacks confinement,
which is consistent with a fixed point
behavior seen in the normalized fourth cumulant of the Polyakov loop.
\end{abstract}
\narrowtext
\section{Introduction}
In some ways, placing a continuum theory on a 
lattice is a dangerous thing.  The 
discreteness of space-time on the lattice 
results in fields which are discontinuous.  If these 
discontinuities are small compared 
to the typical field magnitude, then a reasonable 
interpolation could be defined.  
However, often fields are so discontinuous as to be nearly 
random, and qualitatively different 
interpolating fields could be fit to them.  Such large 
discontinuities can result in 
spurious effects that can exist only in the lattice theory and not 
in the corresponding continuum theory.  
For instance, large amounts of electromagnetic 
flux can be ``lost'' between the links of 
the lattice, creating large violations of Gauss' law.  
This combined with the compact nature of the 
gauge group can result in pointlike or 
stringlike topological defects on the smallest 
1x1 scale: vortices, monopoles, and strings of 
monopole current.  These defects can change the 
qualitative behavior of the theory, for 
instance in the U(1) theory monopoles disorder the 
theory so much that a confinement-deconfinement 
phase transition occurs at a coupling around g = 1, 
resulting in a dramatically different 
new phase on the lattice, a confining one, not connected 
to or relevant 
to the continuum theory.  

The SU(2) theory on a finite lattice also appears to 
undergo a confinement-deconfinement phase 
transition, but this is usually 
interpreted as a finite 
temperature phase transition, one that exists if one of the 
four lattice dimensions is kept 
finite, and the other three are allowed to become infinite.  
This transition is expected to 
disappear in the 4-d symmetric infinite lattice limit.  
This finite-temperature phase 
transition has been linked to the U(1) bulk phase transition 
in the following way.  If the 
SU(2) gauge configurations are transformed into the 
maximum abelian gauge and abelian 
projected to U(1) fields, then the SU(2) finite 
temperature deconfining transition is 
coincident with the monopole-induced transition for the 
corresponding U(1) fields\cite{klsw,ekms}.  
In the confining phase there are large loops 
that tend to span the lattice, perhaps in a 
percolating cluster.  When large loops are not present,
the Polyakov loop shows deconfinement\cite{bmmp}.
In addition, the monopole part of the U(1) 
field seems to carry most if not all of 
the SU(2) string tension\cite{ekms}.  The same
is also true for the U(1) theory itself\cite{stack}.

Although the monopoles are definitely 
artefacts in the U(1) 
lattice gauge theory, they 
are not necessarily artefacts in the SU(2) theory, 
where it is possible that topological 
objects larger than the 1x1 scale exist (fat monopoles and vortices)
that survive the continuum limit, but which map into 
1x1 scale monopoles and their associated thin Dirac strings 
when the abelian projection is performed\cite{fat}.
However, there are 
also undoubtedly 1x1 scale objects in the SU(2) 
theory which {\em are} artefacts, and which also 
show up in the abelian projection.  Thus it would 
seem important to find a way to 
eliminate or at least suppress these SU(2) artefacts, so the 
effects of the large topological 
objects relevant to the continuum limit can be seen.  
The remaining theory, with artefacts 
eliminated, could be substantially different.  With 
less ``noise'' at the smallest scale, it may 
be possible to identify the so far elusive key 
properties of gauge configurations which are 
responsible for confinement.  

\section{Restricted Action}

The aim of this research is to study the effects of 
suppressing artefacts through 
ever greater restrictions on the action, beyond the 
positive-plaquette restriction.  In SU(2) 
and U(1) lattice gauge theory the usual Wilson action 
can be written as $1-P$ where $P$ is the 
plaquette variable which ranges from -1 to 1.  Thus 
restricting the action to be less than a 
certain value is equivalent to restricting the plaquette to 
be greater than some cutoff value.  
The action to be used is the usual Wilson action with the 
added constraint that $P \ge c$, 
where $c$ is some cutoff value.  
Since the continuum limit is determined only 
by the behavior of the 
action in an infinitesimal region around its minimum, which 
occurs around $P=1$,
this action should have the same continuum 
limit as the Wilson action
for all cutoffs $c < 1$.
The case $c = 0$ has been studied before 
as the positive 
plaquette action.  The positive plaquette action eliminates a 
class of SU(2) artefacts, Z2 
monopoles and vortices, which were once thought to possibly 
be the cause of 
confinement.  Mack and Pietarinen found a much smaller and 
non-scaling string tension 
than in the standard action\cite{mp}.  However the theory still 
confined, as confirmed later by 
Fingberg et. al. \cite{fing}.  A similar action, the 
logarithmic action 
was introduced in \cite{log}.  It was 
also shown to confine at very strong couplings\cite{hellog}.

The efficacy of action restrictions in suppressing 
artefacts is demonstrated by the 
U(1) theory, for which a restriction $c > 0.5$ eliminates all 
monopoles.  This can be seen as 
follows.  In  a monopole, 2$\pi$ units of flux enter an 
elementary cube from a Dirac string.  
The flux splits up and emanates in all directions.  Since 
the entering Dirac string is not 
``visible'', this cube looks like a point charge. 
The flux coming out of the 
six plaquettes bounding the 
monopole must add up to the original $2 \pi$.  Therefore, if the 
flux through each plaquette is 
forced to be less than $\pi /3$,
through a plaquette constraint, such a monopole can not exist.  
Since $\cos(\pi /3) = 0.5 $ this 
corresponds to $c > 0.5$ .  Monopoles have also been eliminated 
from U(1) by preventing 
the formation of strings \cite{cahill}.  The action 
restriction idea has 
also been applied successfully 
to spin theories to eliminate
the effects of large discontinuities\cite{ps1}.

The above success in eliminating artefacts suggests 
trying restrictions of order $c=0.5$
in the SU(2) theory as well.  This would likely 
eliminate or at least suppress similar 
non-abelian objects, while allowing larger objects of 
any kind, such as fat monopoles, to still exist.  For example, 
even with a restriction $c=0.5$, the 2x2 Wilson 
loop can still take on any value.  From this 
point of view such a restriction is not very severe.  One 
is simply requiring each plaquette 
to carry less than 1/3 its maximum flux, which is a 
larger amount than could be carried in 
an unrestricted plaquette on a lattice with half the lattice spacing.

\section{Deconfining Phase transition}
To give the restricted action lattices a maximum chance 
to confine, Monte Carlo 
simulations were performed in the strong coupling limit
$\beta \rightarrow 0$,
i.e. the 
configurations are unweighted except that they obey 
the action restriction constraint.  
At least 1000 equilibration sweeps were performed, followed by
from 10,000 
(for the $20^4$ lattices) to 1,500,000 (for some $6^4$ lattices)
measurement sweeps.
Simulations were performed on a large number of Pentium PC's.

As the 
cutoff is raised, a deconfining phase transition occurs
on all lattices studied.  There is 
a fairly strong finite lattice size dependence, with 
the apparent
``critical cutoff'' at around 
0.16 for the $6^4$ lattice, 
0.30 for the $12^4$ and 0.37 for the $20^4$
(these each have an uncertainty of about 0.01).  
It is therefore very important 
to determine the nature of the infinite lattice-size limit.  
If the transition is akin to a finite 
temperature transition, it will be forced to $c = 1$ as the 
lattice size goes to infinity.  On the 
other hand if it is a 4-d percolation transition similar 
to U(1) it will approach a limiting value 
of $c$ which is less than unity, perhaps around 0.5.  Percolation 
transitions can have  
substantial finite size dependences\cite{perc}, so 
this could easily be 
confused with a finite temperature transition.

The behavior
of the Polyakov loop shows the normal symmetry breaking 
behavior and
is smooth, suggesting 
a continuous transition (Fig. 1).
The lower curves in Fig.~1 are the modulus of the spatially
averaged Polyakov loop, $<|L|>$, with the first (absolute value) 
moment of a Gaussian
of the {\em same width} subtracted from it. This gives a  sharper
picture of the phase
transition by correcting for the use of the  absolute value of the
Polyakov loop as the order parameter.  
In the confining phase, where the Polyakov loop 
distribution is very 
close to Gaussian, this subtracted Polyakov loop
is zero within errors, 
whereas it is nonzero in the symmetry breaking region.
Histograms show typical symmetry-breaking behavior
of a higher order transition (Fig. 2).

The normalized fourth cumulant of the Polyakov loop, 
$g_4 \equiv 3 - \!\! <\!\! L^4 \!\! >\!\! /\!\! 
< \!\! L^2 \!\! >^2  $, shows fixed point behavior 
(no discernible lattice size dependence) for $c > 0.5$ 
at a non-trivial value around $g_4 = 1.6$ (Fig 3). 
(The data for $c=0.5$ on the $16^4$ and $20^4$ lattices are
inconclusive as to whether dropping or not).
This suggests either a line of critical points for $c > 0.5$, 
e.g. from a massless gluon phase, or that correlation 
lengths are so large that finite lattice size dependence is hidden. 
For the standard picture of all cutoffs being ultimately 
confining at zero temperature to hold, 
the normalized fourth cumulant should 
go to zero as lattice size approaches infinity for all 
values of $c<1$. (Conversely, above a phase transition it should
go to 2, and at a critical point some non-trivial value in between).  
The susceptibility also appears to diverge with lattice size
for $c>0.5$ \cite{lattice96}, a further indication of criticality
in this region. 

Extrapolations of finite lattice ``critical cutoffs'' to 
infinite lattice size can also be attempted (Fig.~4).
The critical cutoff can be defined many ways, and will
have somewhat different values depending on the definition,
because, after all, the finite lattice system is not really critical
but just showing a rapid change in behavior.
The method used for Fig.~4 was to extrapolate the subtracted
Polyakov loop (defined above) to zero. This quantity 
is consistent with zero in the confining region, and
rises above in the deconfined region.  Quadratic fits, 
which fit the data well in the region just above 
criticality were used. The finite-lattice critical point, $c^*$,
was defined as the point at which the extrapolation hits zero.
Other definitions of finite lattice ``critical cutoffs'' give very
similar looking graphs\cite{lattice96}.
The finite lattice data can then be extrapolated to infinite
lattice size ($N\rightarrow \infty$).
A straightforward linear fit (excluding the $6^4$ point)
gives an infinite lattice critical cutoff of $c^{*}_{\infty}=0.48$.
A fit of the form $c^{*} = c^{*}_{\infty} + c_{1} N^{-0.8}$ can fit
all of the data and gives
$c^{*}_{\infty}=0.49$.
However, it is possible that the graph will curve up sharply
when extremely large lattices are encountered.
As seen from the figure, a logarithmic scaling
function of the form 
$c^* = 1 + a/\ln(b/N)$ can fit the data 
and has $c^{*}_{\infty} = 1.0$ in 
which case the transition
would no longer exist on the infinite lattice. 
Therefore these data alone
do not constitute a definitive test 
of the infinite lattice behavior.
What is needed is a quantity that shows less 
finite lattice dependence,
so that a more reliable extrapolation to the infinite 
lattice can be made.

\section{Monopole Loop Distribution}

The correlation of confinement with the 
appearance of large monopole loops
suggests another more definitive 
approach to extrapolate to the 
infinite lattice. This concerns
the loop size distribution function, 
which for this action appears to follow
a simple power law, as it also appears to do  
for the Wilson action\cite{teper}. 
The power can be extracted from the behavior of small
and mid-sized loops on finite lattices and appears 
to be independent of the 
lattice size. Once the power is known, then 
the probability of having loops
of order the lattice size on lattices of 
arbitrary size can easily be 
predicted. Except for one limiting case, this 
probability will either
vanish or diverge as the lattice size is 
taken to infinity, producing
either a presumably deconfined or confined theory.

Gauge configurations from the restricted action 
simulations were transformed
to maximum abelian gauge using the adjoint 
field method \cite{haymaker}. It was found that this
worked optimally when the adjoint field was
recalculated after each sweep of the gauge field.
Abelian monopole currents were then extracted in 
the usual way using the DeGrand-Toussaint
procedure\cite{dg}. 
Sample sizes ranged from $500$ configurations 
for some $20^4$ lattices, to $200,000$ for some 
$6^4$ lattices.
The imposition of the cutoff 
produces a rather severe
suppression of monopoles, the density of which is 
shown in Fig.~5. The data
are consistent with an exponential suppression of 
the form $\rho \propto
\exp (-k/(1-c))$ as shown by the fits in the figure. 
The $12^4$ fit gives
$k=15.8 \pm 0.3$. A moderate finite-size
shift is seen for the $6^4$ data, but
there is not much difference between the $12^4$
and $20^4$ data. If this exponential continues
for larger $c$, then 
some monopoles will exist for
any value of $c$, making it possible for some to 
survive the continuum limit.

For $c>0.5$ most lattices of practical size have no monopoles, 
e.g. at $c=0.53$
only about one out of every 1000  $12^4$ lattices 
has {\em any} monopoles, usually
a single minimal loop of size four.  Of course, 
even with this low
density, the infinite lattice will still have an 
infinite number of monopoles.
However, what is important is whether they form into large 
loops, because only
these configurations can disorder large Wilson 
loops to produce confinement.
Small loops, such as the most common minimal loop of size four,
will have zero physical size in the continuum limit and presumably
no effect on physics.
Whether large loops exist on large lattices
depends on how fast the 
probability of finding 
loops of size $l$ (i.e. length $l$) 
decreases with $l$.  Define the 
loop distribution function,
$p(l)$, as
the probability
(normalized per lattice site) of finding a monopole loop 
of size $l$ on 
a lattice (of any size). 
Evidence will be presented that $p(l)$ is 
independent of lattice size, for $l$ less 
than several times the 
linear lattice size, $N$.
The probability of finding a loop 
of size $N$ {\em or larger} on an 
$N^4$ lattice is given by $N^4 I(N)$ where $I(N)$ 
is the integrated loop distribution
function
\begin{equation} I(N)=\int_{N}^{\infty} p(l) dl
\end{equation}
where, since $N$ will be taken large, the 
discrete distribution has been
replaced by a continuous one.
To get confinement, at least some 
finite fraction of lattices would
have to contain loops of size order $N$ or larger. 
(Some would
argue that loops of size $N^2$ or larger might be 
necessary to get a linear extent
of order $N$, due to the 
crumpled nature of the loops). Conversely, if 
\begin{equation}
\lim_{N\rightarrow  \infty} N^4 I(N) = 0 \end{equation}
then there will be no loops of size $N$ 
{\em or any finite fraction} of $N$ present on the $N^4$ lattice
in the large lattice limit, 
and the lattice will almost certainly be deconfined
(assuming that large monopole 
loops are necessary for confinement).

In Fig.~6, $\log_{10} p(l)$ is 
plotted vs. $\log_{10}(l)$ for various
cutoffs and lattice sizes. 
The data are consistent with a power law, $p(l) \propto l^{-q}$, 
for loops up to around size $l=3N$
(the size 4 loops, which fall slightly below the 
trend are excluded from 
all fits). 
The larger
the lattice the further the power law 
is valid before some deviation
at large $l$. 
Also note
that the $12^4$ and $20^4$ data are virtually 
identical for loops up to size 30 or so.
Linear fits were made for loop sizes in the range 6 to 2N,
or less if the data had run out (only
the $12^4$ fits are shown for clarity).
For larger loop sizes, occasionally zero instances
of a particular size was observed. These cannot be
plotted, but if they are ignored the data will be
biased upward. A running average procedure was used
in this circumstance to properly account for the 
zero observations. For the most part this was beyond the 
region where fits were performed.  For $c=0.51$, the data were
insufficient to give a reasonable two-parameter fit. Instead,
for this case the constant term was predicted from the 
trend observed for the constant terms of
the other fits, and a one-parameter fit was made for the slope.

The deviations from linearity for large 
loops can be easily understood as a finite size effect
coupled with the periodic boundary condition. For 
loops longer than about $2N$,
there is a significant probability of reconnection through 
the boundary.
This makes a would-be large loop terminate earlier 
than it would on 
an infinite lattice. Thus, on a finite lattice there 
will be a deficit of
very large loops, and an {\em excess} of 
mid-size loops due to this
reconnection effect.  Looking at the data e.g. 
for $c=0.30$, it is apparent
that this is indeed happening. 
The linear trend continues further for the $20^4$ lattice
than for the $12^4$. The observed data does fall below the trendline
for very large loops in the sense that zero instances of loops
beyond those plotted occurred. Of course 
these cannot be plotted on the logarithmic
graph, but the consequences can be taken into 
account in the following way. If one assumes that very large loops
follow the trendline
in the figure for the $12^4$ data, 
then one can calculate that 5.3 instances
of loops in the size range 342 to 2000 should have been seen in
the sample. The fact than none occurred implies rather strongly
that the data does eventually fall below the trend line for very
large loops ($p<0.01$).	Similar arguments 
can be applied to the other
data samples.

Because the power 
law trend continues further
the larger the lattice, 
and the deviations are easily understood as a finite size
effect, it seems quite reasonable to assume that on the 
infinite lattice one would
have a pure power law. 
It is difficult to imagine what could set the scale for 
a significant change in behavior at extremely large loop sizes beyond 
those measured here.
In addition, since the small loop data are nearly
independent of lattice size, it would seem the power 
must also be essentially the same
on the infinite lattice as observed here for 
the $12^4$ or $20^4$ lattices.  
Assuming this, one can easily predict the point at 
which condition (2) 
becomes satisfied, namely $q>5$. For $q>5$ the 
probability of having
a monopole loop with length equal 
to any finite fraction of the lattice
size $N$ vanishes in the large lattice limit, 
whereas for $q<5$ the same 
becomes overwhelmingly likely. 

The power $q$ is plotted 
as a function of cutoff in Fig.~7. A definite
rising trend is observed, 
with $q$ passing 5 around $c=0.45$.  It is very
difficult to gather enough statistics for $c>0.5$
since monopoles are extremely rare here, however if all
of our runs in this region are combined(c=0.51 to 0.55), 
then the following statement
can be made. 
Out of a total sample of over 10 billion links, only a single
loop of size 8 was found (at $c=0.51$), 
and none larger, whereas 230 loops of size 4
and 36 of size 6 were found. 
If $q \le 5$ then the expected number of size 8 
loops, given this many size six loops, would be at least
8, and several even larger loops should have been seen.
Using Poisson statistics,
the probability of obtaining our result 
for size 8 and larger loops
if $q \le 5$ can be computed 
to be around $10^{-5}$.
Thus it appears overwhelmingly likely that $q$ exceeds 5
for $c \ge 0.51$. Therefore, the loop distribution function
strongly supports the notion that this theory is deconfined
in the infinite lattice limit for $c>0.5$. This is 
in concert
with the results from the fourth cumulant of the Polyakov 
loop (Fig. 3), showing fixed-point behavior in this range,
and with the straightforward extrapolation of 
critical cutoffs (Fig. 4).

The rather strong finite lattice size dependence 
of the critical cutoff
in this theory, or critical $\beta$ in the standard 
Wilson-action theory
can be understood from the following argument. 
Say that the theory 
becomes confining at the point that 
$N^4 I(N) = 0.5$, i.e. when 50\% 
of the lattices of size $N$ have 
a loop of length at least $N$ (the 
exact criterion is irrelevant). 
In the region $q<5$, the LHS is an
increasing function of $N$, so it will 
be satisfied for some $N$. If
$c$ is raised, then $q$ will increase, requiring 
a larger $N$ to
stay on the transition. If $q$ varies relatively 
slowly with $c$ then
there will be a substantial change in the 
critical value of $c$	as
N is changed, until $q$ gets close to 5, at which
point the critical cutoff will reach its limiting value.
It is very likely that changing $\beta$ 
in the standard
theory has a similar effect to changing $c$. 
To test this some preliminary
runs were performed with the standard Wilson action on a $12^4$
lattice. Contrary to the 
suggestion in \cite{teper}, a rather 
substantial dependence of $q$ on 
$\beta$ is found, with $q \approx 3$ 
at $\beta=2.4$ and $q \approx 5$
at $\beta = 2.9$. These data predict that 
the standard Wilson-action
theory will be deconfined on the infinite lattice 
for all $\beta > 2.9$.
Details will appear in a separate report, when 
greater statistics become
available.

\section{Discussion}

The above results suggest 
that the possibility that the continuum SU(2) pure
gauge theory may not be a 
confining theory needs to be taken seriously.
This has been
suggested before\cite{zp,ps2,ps3}. In fact 
Ref. \cite{ps3} predicts that a sufficiently strong plaquette
restriction would result in a theory that is non-confining for 
all couplings and temperatures.

Of course, one could 
instead give up the link between abelian monopoles and confinement,
despite the overwhelming evidence in favor of a connection.
Since abelian monopoles have been shown to be responsible for most if
not all of the SU(2) string tension, if the theory still confines when
they are removed it will be a more subtle form of confinement, with a 
likely much smaller string tension. 
However	string tension is probably not the best test of whether 
a lattice is confining or not.
The Polyakov loop 
is a much better order parameter for confinement 
simply because it does concern an actual symmetry breaking.
It is very difficult to 
tell if a string tension is exactly zero due to 
other terms in the potential and 
the functional forms assumed for them\cite{lcp}.
There is no problem defining 
the Polyakov loop on a symmetric lattice and, although it
does go to zero in both phases as the lattice size 
$N \rightarrow \infty$, 
one can still 
look for symmetry breaking at any 
large finite N, as large as one likes,
or take the $N^{th}$ root and then 
the limit $N \rightarrow \infty$. Normalized
cumulants such as $g_4$ will also 
have nontrivial $N \rightarrow \infty$
limits that allow one 
to distinguish broken from unbroken symmetry behavior
by looking for non-Gaussian behavior in the limit of large 
lattices.
The case for deconfinement in the restricted 
action theory from the Polyakov loop and its moments alone
is fairly compelling, though not as definite 
as the monopole loop data, due to the large finite size dependence
of the critical cutoff. Nevertheless, it supports the notion that
deconfinement results when large abelian monopole loops disappear.

Our simulations are in the strong coupling limit. Letting $\beta$
grow larger than zero will further order the theory. If the theory
is already deconfined in the strong coupling limit, 
it is very unlikely that confinement could come back as the
coupling is weakened. Thus the zero-temperature continuum limit, 
$\beta \rightarrow \infty$ would also be deconfined.

The behavior of the standard theory in
the fundamental-adjoint plane can also be interpreted 
as supporting this conjecture,
as it suggests that what is normally thought of as a 
finite-temperature transition may actually be a 
zero-temperature bulk 
transition, since it appears to connect to a 
previously known bulk transition\cite{ggm}.
Although some evidence of a separation of the 
bulk and finite temperature
transitions has been presented\cite{fa}, 
this can be interpreted merely as a 
manifestation of the fact that different methods 
of finding a critical
point on a finite lattice will usually give 
slightly different values,
agreeing only in the thermodynamic limit. There 
has yet to be a simulation
showing two distinct transitions at different 
$\beta$'s on the same lattice.
By bulk transition, 
it is meant here a transition that remains at finite
$\beta$ in the infinite (4-d) lattice limit 
(as opposed to a finite-temperature
transition for 
which $\beta_c \rightarrow \infty$ in this limit).
If the deconfinement transition
is a percolation transition similar to U(1) this will be true. However
percolation transitions differ in one respect
from what is usually called a bulk transition
in that only a fractal
network of links comprising a small fraction of the 
total set of lattice links actually
participates in the transition, so scaling properties will 
likely differ from 
a conventional bulk transition in which all plaquettes 
participate.  This 
may explain the different from bulk scaling exponents 
seen in \cite{rajiv2}
for the first-order transition seen in the case of a large 
adjoint action.
It could also explain the larger than normal finite-size shift 
in critical point for the standard Wilson action theory, since 
this shift
is related to the scaling exponents.

It is important to ask the question of whether or 
how continuum QCD could 
live with a non-confining continuum SU(2) theory.  
First, the behavior of SU(3) could differ. Although this certainly
needs to be checked, there has always been a qualitative agreement
between these theories so far.
Another possibility is that confinement is not absolute, in the 
sense of
a linearly rising potential that goes on forever.
All that is needed in the real world
is for the potential to have a nearly linear portion in the range
1-5 fm.  Beyond this, particle pair creation causes the ``string''
to break in the real world, so details of the potential 
at larger distances in the pure gauge theory are 
irrelevant to experiment. A logarithmic
running coupling can modify the Coulomb potential to produce a 
potential that is nearly linear in this range, but at large
distances
goes to a constant \cite{lcp}. 
This may be enough to fit heavy quark spectroscopy.

Another possibility is that confinement 
could be due to chiral symmetry breaking\cite{zp,nc}.  
With light fermions present it is likely that chiral symmetry
will still break in a non-confining theory, since the coupling is
strong. Confinement could then result from a polarization of
the chiral vacuum which results in a higher than normal vacuum
energy density in the region surrounding a colored object, including 
color dipoles such as mesons and baryons. This region of polarized
vacuum moves 
around with the meson or baryon adding to its dynamical mass.  
Any attempt to stretch the hadron will stretch this ``disturbed
vacuum'' bag leading to an energy proportional to the elongation,
i.e. a linear potential.  This picture is consistent with
the observation that $<\! \bar{\psi}\psi \! >$ is lowered in the 
neighborhood of a color source\cite{markum}, indicating some 
expulsion of 
condensate. Since the condensate is expelled, the energy
density must be increased, supporting the above picture, 
in which chiral symmetry breaking, confinement, and dynamical 
mass generation of quarks are all due to the same mechanism.
This scenario is similar to the picture that emerges in chiral
quark models\cite{cqm} where a polarized Dirac sea is responsible 
for the binding of the quarks in a baryon, and also in the 
instanton liquid model\cite{ilm}. Both of these models
are able to compute with fair accuracy a large number of 
low-energy properties of hadrons, and neither has an
absolutely confining 
potential.

As a final note, from the point of view of 
practical simulations, it may be better to take 
an action that is cut off more smoothly than the one
considered here. A smooth cutoff action 
that disallows plaquettes smaller than a cutoff $c$ is
\begin{equation}
S_{\Box} = \left\{ \begin{array}{ll} 
-(1-c) \ln \left[
(P-c)/(1-c) \right] & \mbox{if $P>c$}\\
                  \infty  &  \mbox{if $P \le c$}
\end{array}
\right.    \end{equation}
where $P$ is the plaquette.
The smoothly cutoff action may have better scaling 
behavior, as has been shown for
a similar logarithmic action based on the positive 
plaquette action\cite{hellog}.

\section{Conclusion}
The imposition of a plaquette 
restriction causes a deconfining phase
transition in SU(2) lattice gauge theory, 
even in the strong coupling limit.
The critical cutoff is dependent on lattice size. 
Straightforward extrapolation as well as the behavior of
the fourth moment of the Polyakov loop suggest that
the infinite lattice critical cutoff will 
be around $c=0.5$, the same
value for which the U(1) theory must deconfine. 
This means that the theory will be deconfined on all 
symmetric lattices for $c>0.5$, for any $\beta$.
The abelian monopole
loop distribution function confirms this by 
showing a power-law falloff with loop length $l$
faster than $l^{-5}$ for $c>0.5$, from which it can be shown that 
no loops large enough to cause confinement exist on any size symmetric
lattice.  Light dynamical quarks may be a necessary ingredient 
to obtain a continuum confining theory.

\acknowledgments
It is a pleasure to thank 
Richard Haymaker for advice on implementing
the adjoint field method.

\newpage
\begin{center}
{\Large Figure Captions}
\end{center}
\noindent
FIG. 1. Typical modulus of the Polyakov loop.

\noindent
FIG. 2. Polyakov loop histograms for confined and deconfined 
$8^4$ lattices.

\noindent
FIG. 3. Normalized Fourth cumulant of the Polyakov loop. Errors are
from binned fluctuations.

\noindent
FIG. 4. Extrapolation of critical point to infinite volume. 
Uncertainties
are about the size of plotted points. $N$ is the linear lattice
size. The short-dashed line is a linear fit, longer dashed line a
fractional power fit, and the solid line is a fit to 
a logarithmic function given in the text.

\noindent
FIG. 5. Logarithm of the monopole (plus antimonopole) 
density (number per lattice link) vs. $1/(1-c)$. A linear fit to the 
$12^4$ data is also shown.

\noindent
FIG. 6. Log-log plots of loop probability (per lattice site) vs. loop
length. The six data series shown are, from right to left, $c=$ 0.30, 
0.36, 0.42, 0.45, 0.49, and 0.51. Trend lines explained in text are 
given for the $12^4$ data. Only $12^4$ runs were 
performed at the highest
two cutoffs. Size four loops fall below the trend and are 
not included 
in fits.

\noindent
FIG. 7. Power, $q$, describing the falloff of loop 
probability with
loop length, vs. the cutoff, $c$. 
Line is drawn through the $12^4$
data to guide the eye. 

\begin{references}

\bibitem{klsw} A.S. Kronfeld, M.L. Laursen, G. Schierholz, and 
U.J. Weise, Phys. Lett. B {\bf 198}, 516 (1987);
T. Suzuki and I. Yotsuyanagi, Phys. Rev. D {\bf 42}, 
4257  (1990); T. Suzuki, Nucl. Phys. B (Proc. Suppl.) {\bf 30} (1993)
176.
\bibitem{ekms} S. Ejiri, S-i. Kitahara, Y. Matsubara, and T. Suzuki,
Phys. Lett. B {\bf 343}, 304 (1995); J.D. Stack, S.D. Neiman, and
R.J. Wensley, Phys. Rev. D {\bf 50}, 3399 (1994).
\bibitem{bmmp} V.G. Bornyakov, V.K. Mitrjushkin, 
and M. M\"uller-Preussker,
Phys. Lett. B {\bf 284}, 99 (1992); T. Suzuki et. al., Phys. Lett. B
{\bf 347}, 375 (1995).
\bibitem{stack} J.D. Stack and R. Wensley, 
Nucl. Phys. B {\bf 371}, 597 (1992);
Phys. Rev. Lett. {\bf 72}, 21 (1994);
W. Kerler, C. Rebbi, and A. Weber, 
Nucl. Phys. B (Proc. Suppl.) {\bf 47}, 667 (1996).
\bibitem{fat} T.L. Ivanenko et. al., Phys. Lett. B {\bf 252},
631 (1990); T.L. Ivanenko, A.V. Pochinsky, and M.I. Polikarpov,
Nucl. Phys. B (Proc. Suppl.) {\bf 30}, 565 (1993);
Phys. Lett B {\bf 302}, 458 (1993).
In a somewhat different context, fat SO(3) monopoles 
are discussed in
T.G. Kov\'{a}cs and E.T. Tomboulis, report 
UCLA/97/TEP/22 (hep-lat/9711009);
Nucl. Phys. B (Proc. Suppl.) {\bf 53}, 509 (1997). 
\bibitem{mp} G. Mack and E. Pietarinen, 
Nucl. Phys. {\bf B205 [FS5]}, 141 (1982).
For a more recent measurement see K. Cahill and G. Herling,
report NMCPP/97-17 (hep-lat/9801009).
\bibitem{fing}J. Fingberg, U.M. Heller, and V. Mitrjushkin,
Nucl. Phys. B {\bf 435}, 311 (1995).
\bibitem{log} M. Grady, Phys. Rev. D 42 (1990) 1223.
\bibitem{hellog}U.M. Heller, Nucl. Phys. B {\bf 451}, 469 (1995).
\bibitem{cahill} K. Cahill and G. Herling, 
Nucl. Phys. B (Proc. Suppl.) {\bf 53}, 797 (1997).
\bibitem{ps1}A. Patrascioiu and E. Seiler, Nucl. Phys. B {\bf 305},
623 (1988); J. Stat. Phys. {\bf 69}, 59 (1992); 
Nucl. Phys. B (Proc. Suppl.) {\bf 30}, 184 (1993).
\bibitem{perc} K. Binder and D.W. Heermann, {\bf Monte 
Carlo Simulation in Statistical Physics}, 2$^{nd}$ corrected
ed., Berlin, springer-Verlag, 1992 pp. 35-41; D. Stauffer
and A. Aharony, {\bf Introduction to Percolation Theory},
$2^{nd}$ revised ed., London, Taylor and Francis, 1994,
pp. 70-88.
\bibitem{lattice96} M. Grady, Nucl. Phys. B (Proc. Suppl.)
{\bf 53}, 599 (1997).
\bibitem{teper}A. Hart and M. Teper, Nucl. Phys. B 
(Proc. Suppl.) {\bf 53}, 497 (1997) (hep-lat/9606022);
Nucl. Phys. B (Proc. Suppl.) 
to appear (Lattice 97) (hep-lat/9709009).
\bibitem{haymaker}K. Bernstein, G. Di Cecio, and R. W. Haymaker,
Phys. Rev. D {\bf 55}, 6730 (1997).
\bibitem{dg} T.A. DeGrand and T. Toussaint, Phys. 
Rev. D {\bf 22}, 2478 (1980). 
\bibitem{zp}M. Grady, Z. Phys C, {\bf 39}, 125 (1988).
\bibitem{ps2}A. Patrascioiu, E. Seiler, and I.O. Stamatescu,
Nuovo Cimento D {\bf 11}, 1165 (1989).
\bibitem{ps3} A. Patrascioiu, E. Seiler, V. Linke, and
I.O. Stamatescu, Nuovo Cim. {\bf 104B}, 229 (1989).
\bibitem{lcp}M. Grady, Phys. Rev. D {\bf 50}, 6009 (1994).
\bibitem{ggm} R.V. Gavai, M. Grady, and M. Mathur, 
Nucl. Phys. B {\bf 423}, 123 (1994).
\bibitem{fa}T. Blum et. al., Nucl. Phys. B {\bf 442} 301 (1995).
\bibitem{rajiv2}R.V. Gavai, Nucl. Phys. B {\bf 474}, 446 (1996). 
\bibitem{nc} M. Grady, Nuovo Cim. {\bf 105A}, 1065 (1992).
\bibitem{markum}W. Feilmair, M. Faber, and H. Markum, Phys. Rev. D,
{\bf 39}, 1409 (1989).
\bibitem{cqm} Chr.V. Christov et. al., Prog. Nucl. 
Part. Phys. {\bf 37}, 91 (1996) and references therein.
\bibitem{ilm}E.V. Shuryak, 
Nuclear Phys. B {\bf 203}, 93, 116, 140
(1982), {\bf 214}, 237 (1983); D.I. Diakonov and V. Yu. Petrov,
Nucl. Phys. B {\bf 245}, 259 (1984), {\bf 272}, 457 (1986).

\end{references}
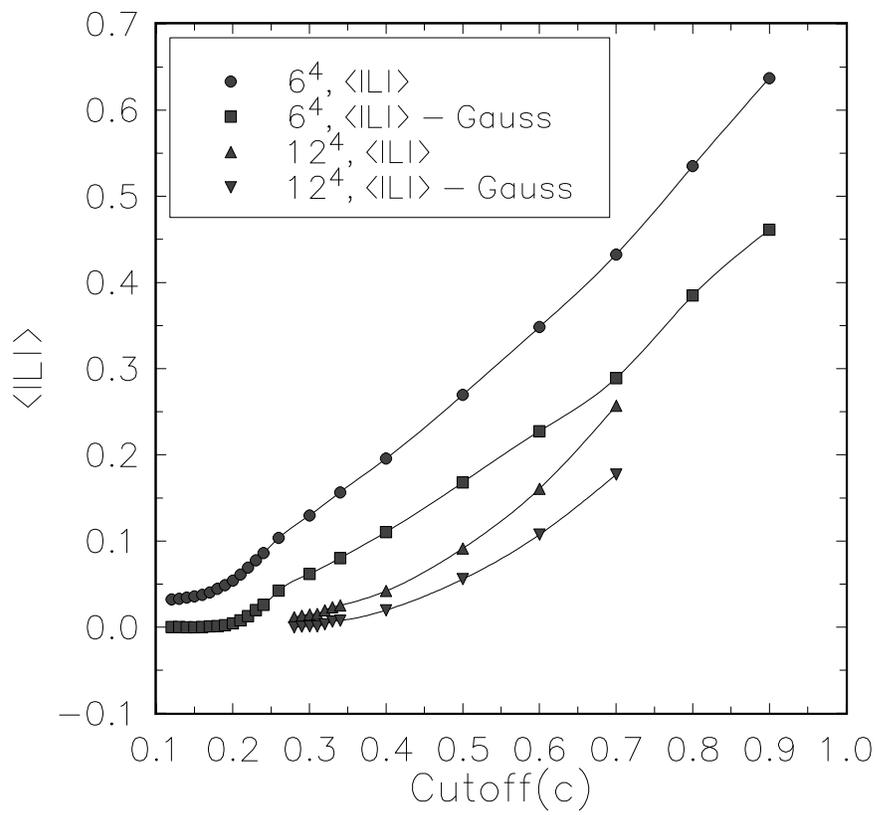
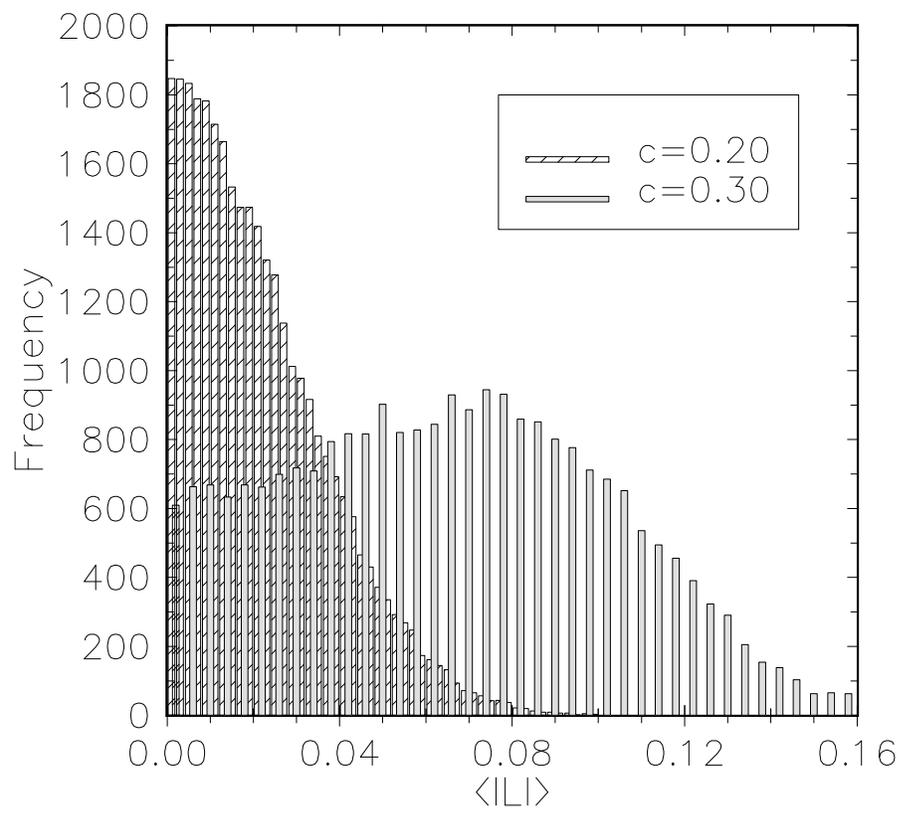
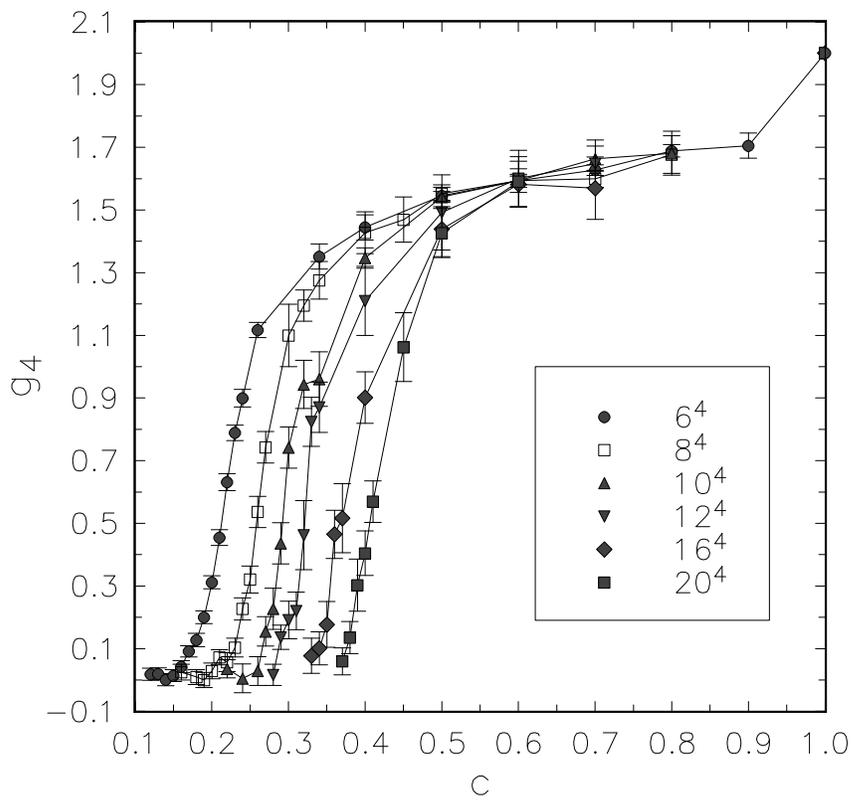
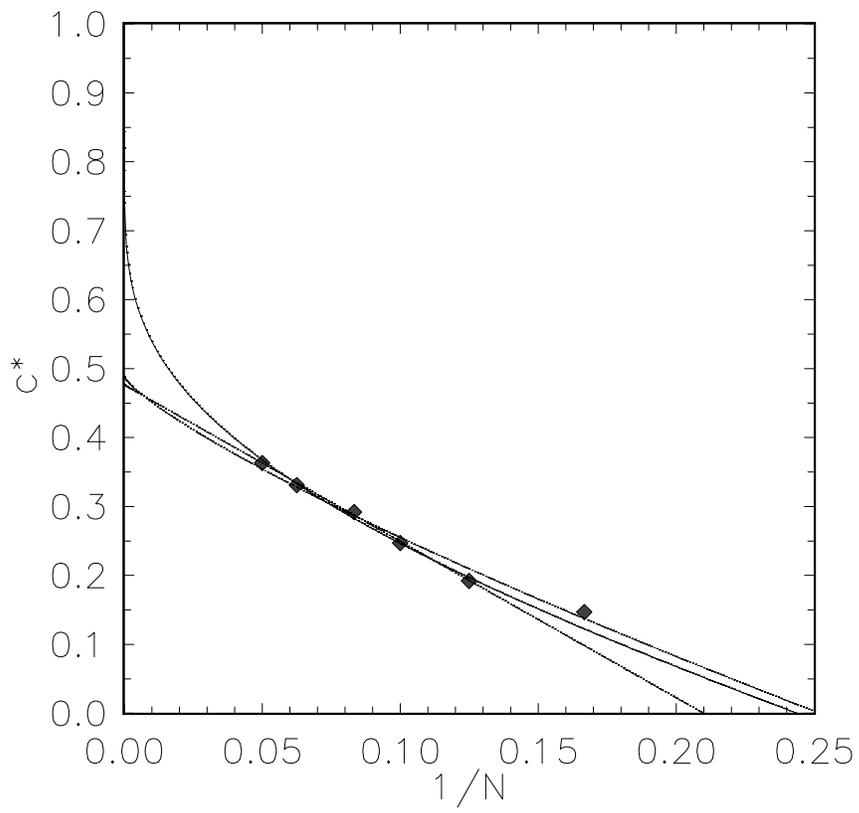
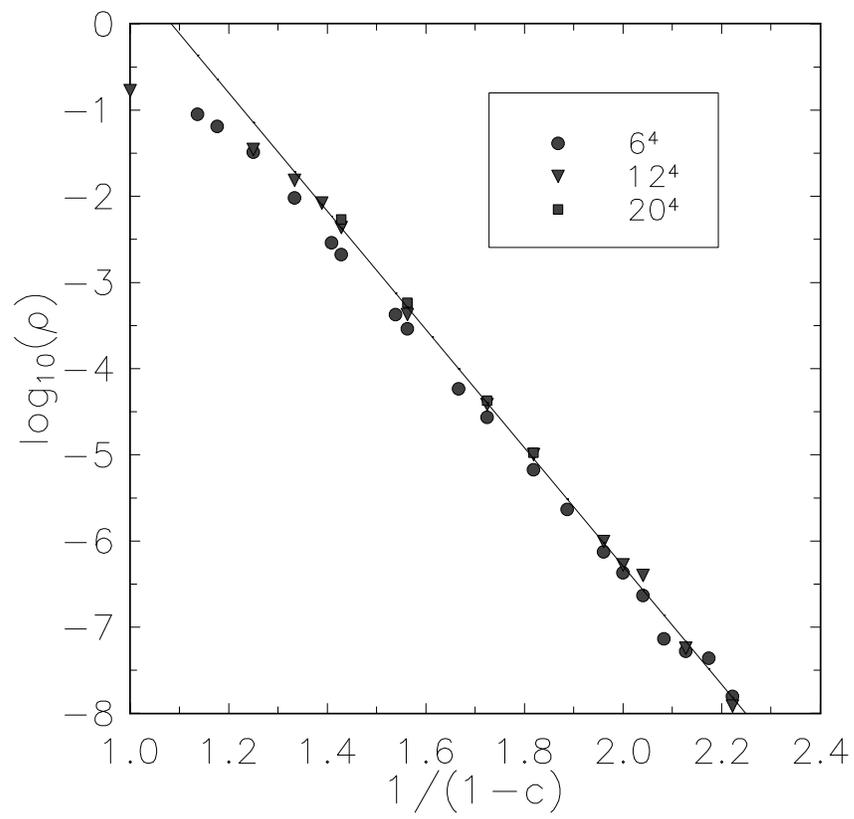
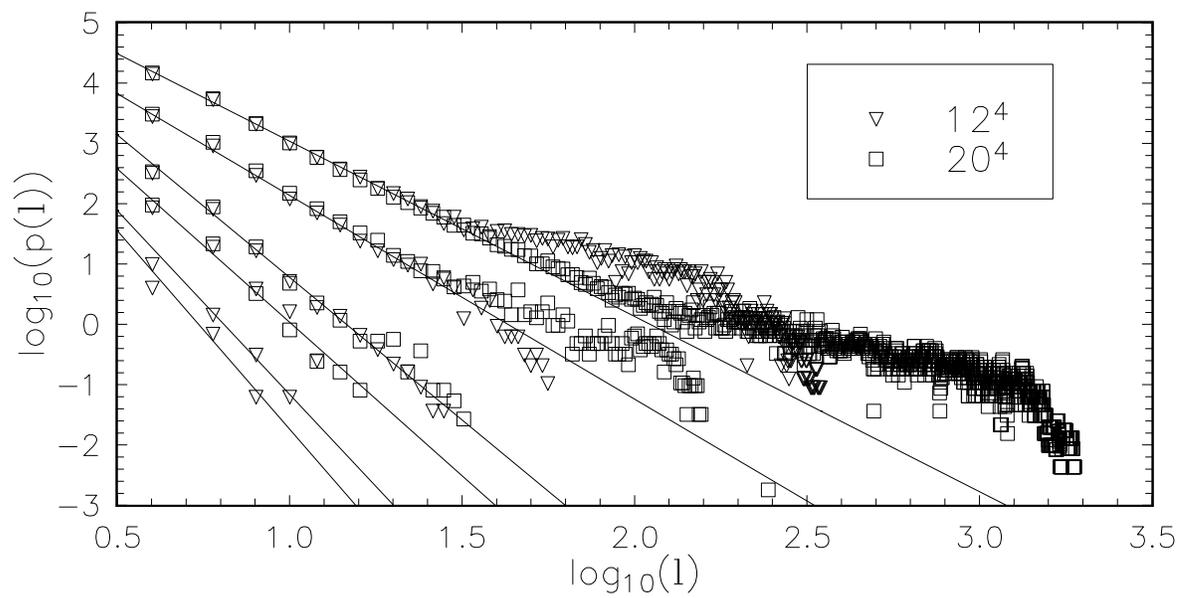
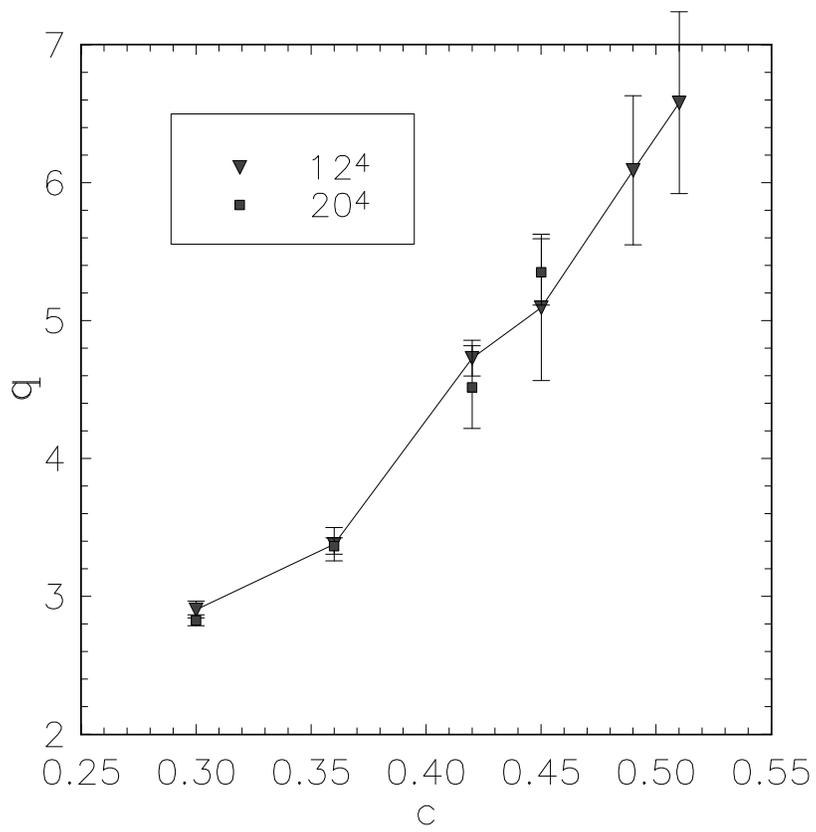
\end{document}